\documentclass[aps,prb]{revtex4-2}

\usepackage[english]{babel}

\usepackage{amsmath,eqnarray,braket,graphicx,csquotes,cases,physics,mathtools,amssymb,siunitx,comment}
\usepackage[colorlinks=true, allcolors=black]{hyperref}
\usepackage{lmodern}
\usepackage{textcomp}
\usepackage{multirow}
\usepackage{dcolumn}
\usepackage{verbatim}
\usepackage{amsmath}

\newcommand{\be}{\begin{equation}}
\newcommand{\ee}{\end{equation}}
\newcommand{\bea}{\begin{eqnarray}}
\newcommand{\eea}{\end{eqnarray}}

\usepackage[dvipsnames]{xcolor}

\begin{document}
\title{Signatures of Chiral Phonons in MnPS$_3$ from first principles}
\author{Banhi Chatterjee and Peter Kratzer}
\affiliation{Fakult{\"a}t f{\"u}r Physik and CENIDE, Universit{\"a}t Duisburg-Essen, Lotharstr. 1, 47057 Duisburg, Germany}
\date{\today}
\begin{abstract}
Two-dimensional (2D) materials may host circular phonons, considered as chiral if the presence of a substrate breaks mirror symmetry. 
In 2D transition metal dichalcogenide (TMDC) monolayers lacking inversion symmetry, phonons with a given chirality  can be observed in the non-equilibrium state triggered by optical excitations using circularly polarized light. 
Backed by first-principles calculations, we present the antiferromagnetic semiconductor MnPS$_3$ with a hexagonal crystal structure and bandstructure similar to TMDCs, but a larger unit cell, as a novel candidate material that may allow for excitation of circular phonons. 
Using DFT+U and the finite displacement method we obtain in-plane chiral phonon modes at the valley points of a monolayer MnPS$_3$. These modes can be classified according to the Mn or S atoms performing circular motions about their equilibrium positions. In each case, the quantized angular momentum of the phonons is calculated. 
Moreover, we point out ways to populate the chiral phonons selectively via optical excitation with circularly polarized light.
\end{abstract}
\maketitle
\section{Introduction}
\label{sec:intro}
Textbook knowledge tells us that 
the elementary excitations of a crystalline lattice give rise to quasi-particles known as phonons; yet some of their properties have remained elusive until recently.
While the analog to linear momentum for phonons, the crystal momentum $\mathbf{k}$, has long been known, the {\em angular} momentum of a phonon is still subject to debate. 
As a finite crystal, seen as a rigid body, can have a well-defined total angular momentum, one would expect this property to carry over its smaller constituents, and ultimately to the atoms in a unit cell of the crystal. 
These considerations must play a role in the microscopic understanding of the Einstein-de Haas effect~\cite{zhang2014Einstein,tauchert2022baum} where the rigid-body rotation builds up gradually during the quench of magnetism, or more generally in any situation where spin or orbital angular momentum of the electronic system are transferred to the crystal lattice. Conversely, orbital magnetism may arise from circular motions of the ions in a crystal~\cite{chaudhary2024giant,chaudhary2025anomalous,juraschek2022giant}. 
Therefore, investigating angular momentum transfer in a crystal is a meaningful and interesting research question.

Since a crystal, unlike free space, is invariant under rotations only for specific axes and a discrete set of rotation angles, it follows that the angular momentum of a phonon will not be a conserved quantity in general. 
However, there are specific situations, particularly lattices displaying a three-fold rotational symmetry axis, e.g. hexagonal, honeycomb or Kagom\'e lattices. 
If we label the rotational axis the $z$-axis, it is no longer meaningful in these crystals (unlike e.g. in cubic crystals) to decompose the atomic displacements in the $xy$-plane into separate $x$ and $y$-modes. Rather, the symmetry-adapted motions are superpositions in which the atoms run on circles or ellipses in the $xy$-plane. We speak of these modes as circular phonons. For circular motion around the three-fold axis, quantization leads to phonons having an angular momentum of $\pm \hbar/3$ \cite{zhang2015chiral}. 
This pseudo-angular momentum (PAM) is a conserved quantity. 
The conservation law for PAM makes it possible to consider angular momentum transfer from elliptically (or, more specifically, circularly) polarized light via the electronic subsystem to the lattice, thereby temporarily populating only phonon states with a specific sense of rotation (a specific sign of the PAM) under non-equilibrium conditions. 

Circular phonons have been studied computationally in various two-dimensional materials, such as graphene or transition-metal dichalcogenides (TMDCs)~\cite{zhang2015chiral}.
If placed on a substrate, the combined system lacks mirror  symmetry, with the effect that clockwise and counterclockwise circular motions cannot be mapped onto each other. In this case, one can speak of chiral phonons.

We note in bypassing that a different, propagating type of chiral phonons can be observed in 3D crystals having a screw axis and thus belonging to a non-symmorphic space group, e.g. $\alpha$-quartz or $\alpha$-HgS \cite{ueda2023chiral,ishito2023truly}. This topic is outside the scope of this paper. 

If phonons of a given chirality should be excited optically by elliptically polarized light, it is crucial that the material lacks inversion symmetry; otherwise chirality-selective coupling of the light to the electronic system will not be possible. Monolayer WSe$_2$ (in contrast to bilayers) meets this condition, and indeed chiral phonons have been optically detected in this material \cite{zhu2018observation}. 

Chiral phonons have also been theoretically studied in the non-magnetic monolayers of MoS$_2$ \cite{pan2023vibrational} and strained WS$_2$ \cite{pan2024strain-induced} which have a hexagonal lattice structure with broken inversion symmetry. Spin-orbit coupling leads to band splitting and valley-polarized population can be selectively excited at the $K^{+}$ or $K^{-}$ points of the Brillouin zone depending on light helicity. After relaxation by electron-phonon coupling, the resulting vibrational state is characterized by a distinct chirality. 
In this paper, we introduce MnPS$_3$, a direct semiconductor, as another candidate to study chiral phonons. Its antiferromagnetism breaks inversion symmetry of the electronic system, while the lattice still possesses an inversion center. 
With a band topology of the valence and conduction bands similar to TMDCs (e.g. MoS$_2$), coupling of angular momentum into the electronic system via elliptically polarized light \cite{li2013coupling} is a viable option to populate phonons of one chirality only. However, in comparison to MoS$_2$, the unit cell of MnPS$_3$ is larger, 
allowing for a richer set of chiral modes. 
In addition to the somewhat indirect optical techniques, structural probes such as X-ray or electron diffraction could be employed to provide evidence for chiral phonon population. Exemplarily, temporarily induced dichroism in the inelastic diffraction signal \cite{britt2023ultrafast} could give a hint, provided the chiral phonon amplitude is sufficiently strong. 
Due to large (but compensating)  magnetic moments at the Mn atoms, orbital momentum arises naturally in the electronic system in the presence of spin-orbit coupling. Thus it appears plausible that this electronic angular momentum can be passed on to the lattice via electron-phonon coupling. The magnon-phonon coupling experimentally observed in this material \cite{matthiesen2023controlling} points to a considerable coupling between magnetic and lattice degrees of freedom. 
This perspective spurs additional interest into studying MnPS$_3$ and related transition-metal trisulfides. \\
In this paper we compute the phonon dispersion in  MnPS$_3$ using DFT+U and the fixed displacement method. We further compute the phonon circular polarization following the theoretical framework presented in Ref.~\onlinecite{zhang2015chiral} and obtain signatures of chiral phonon modes.


 \begin{figure} [ht]
 \centering
 \includegraphics[width=0.8\textwidth]{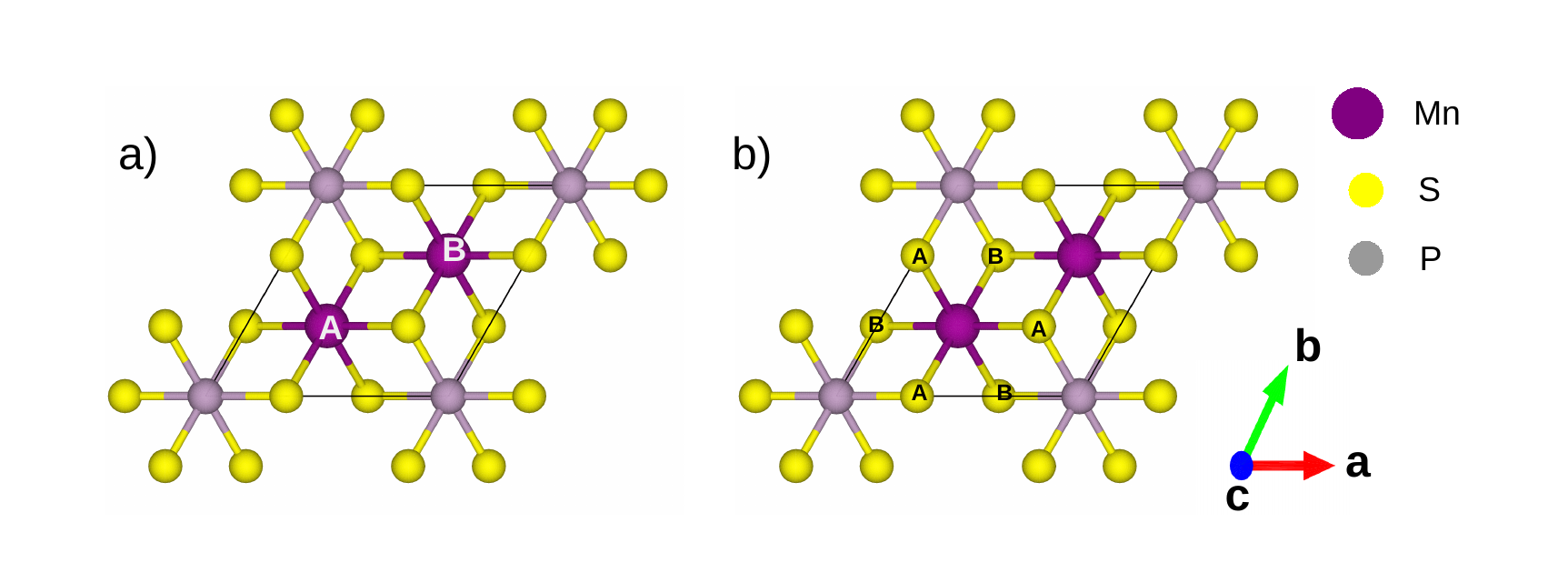} 
 \caption{Crystal structure of a monolayer MnPS$_3$ in anti-ferromagnetic phase with a) the two non-equivalent Mn atoms as site A and site B, b) the S atoms as site A and B as marked. We study the monolayer in a-b plane.}
 \label{structmnps3}
 \end{figure}
 \section{Theory}
\label{sec:matmethod}

Characterizing the chirality of a phonon requires some mathematical background. 
As a starting point, we define the atomic positions $\mathbf{r}_{n\alpha}(t)$ of the atoms indexed by $\alpha$ in the unit cell with index $n$ of a crystal as
\be
\mathbf{r}_{n\alpha}(t) = \mathbf{R}_{n\alpha} + \mathbf{u}_{m\alpha}(t)
\ee
The time-dependent displacements $\mathbf{u}_{n\alpha}(t)$ are used to define the angular momentum $\mathbf{J}^{ph}$ of a phonon
\be
\mathbf{J}^{ph}=\sum_{n\alpha} m_{\alpha} \mathbf{u}_{n\alpha}\times \dot{\mathbf{u}}_{n\alpha}
\ee
where $m_{\alpha}$ is the mass of species $\alpha$. 
In a two-dimensional situation, this simplifies to
\be
 J^{ph}_{z}= \sum_{n\alpha} m_{\alpha} (u^{x}_{n\alpha}\dot{u}^{y}_{n\alpha}-u^{x}_{n\alpha}\dot{u}^{y}_{n\alpha})
\ee
Second quantization and Bloch's theorem in the periodic crystal lead to
\be
\mathbf{u}^{j}_{n,\alpha}= \sum_{\mathbf{k},\sigma} \mathbf{\epsilon}_{\alpha, j} (\mathbf{k},\sigma) e^{i(\mathbf{R}_{n}\cdot \mathbf{k} -\omega_{k}t)} c_k \hat a_k + \mbox{h.c.}
\label{eq:2nd_quanti}
\ee
with the creation and annihilation operators $\hat a^{\dag}$ and $\hat a$, and $c_k=\sqrt{\frac{\hbar}{2\omega_{k} N m\alpha} }$ a normalizing constant. 
Here, $\mathbf{k}$ is the crystal momentum and $\omega_k$ the mode frequency, while $\sigma$ indexes the phonon branch. 

In general, one has to distinguish between the symmetry representations of the crystal lattice itself, and the representation of a phonon. 
Given a hexagonal crystal with 3-fold rotational symmetry, the phonon representation shares this symmetry if we restrict ourselves to the $\mathbf{k}$ vectors at high-symmetry points in the Brillouin zone. 
At the $\Gamma$ point, the phonon representations are identical to the lattice representation. 
The more interesting case occurs for $\mathbf{k}=K^+$ or $\mathbf{k} =K^-$, i.e., at one of the six corners of the Brillouin zone that are decomposed into the two subsets $K^+$ and $K^-$. These two subsets are mapped onto each other by inversion, $\mathbf{k} \mapsto -\mathbf{k}$. 
Due to the lack of inversion symmetry on both the TMDC and MnPS$_3$ monolayers, it is essential to distinguish  $K^+$ and $K^-$.
The 3-fold rotational symmetry at these points leads to phonons that either transform according to a one-dimensional $A$ representation or two-dimensional $E$ and $E^*$ representations. 
To exemplify this, let ${\cal R}_z(\frac{2 \pi}{3})$ denote a rotation around the $z$-axis by $120^{\circ}$.
According to the $E$ or $E^*$ representation, the Bloch factor in Eq.~\ref{eq:2nd_quanti} is mapped to
\be
 {\cal R}_z \left(\frac{2\pi}{3} \right) e^{i \mathbf{R}_{n} \cdot \mathbf{k}  } = e^{i l^o 2 \pi/3} e^{i \mathbf{R}_{n} \cdot \mathbf{k}  }
\ee
This introduces the quantum number $l^o = \pm1$, describing a pseudo-angular momentum quantized in units of $\pm \hbar/3$. 
Since it originates from the longe-range behavior of the displacement pattern, it has been dubbed the "orbital" angular momentum of the phonon, in loose analogy to the terminology in magnetism.

In unit cells with several atoms, the question arises what is the contribution of the displacement pattern inside the unit cell, encoded in the eigenvector $\mathbf{\epsilon}(\mathbf{k},\sigma)$, to the angular momentum. This microscopic contribution has been termed the "spin" angular momentum of the phonon. To define the "spin", we start from the phonon circular polarization (PCP)
\be
s_{z}^{\sigma} =i\sum_{\alpha}[\mathbf{\epsilon}^{\dag}_{\alpha,y} (k,\sigma) \mathbf{\epsilon}_{\alpha,x} (k,\sigma) - \epsilon^{\dag}_{\alpha,x} (k,\sigma) \epsilon_{\alpha,y} (k,\sigma)]
 \label{pcp}
\ee
The overall angular momentum of the phonon is given by the sum of both, orbital and spin, angular momentum,
$l^{ph} = l^o + l^s$. It has been argued \cite{zhang2015chiral} that the spin angular momentum at $K^+$ or $K^-$ needs to be quantized in units of $\pm \hbar /3$, too. 
The three-fold rotational axis may run through any selected atom in the unit cell. Either this atom is standing still, or it must move, clockwise or counterclockwise,  on a circular trajectory; no other displacement pattern is possible under these symmetry conditions. 
Different sublattices of the unit cell, or more generally distinct lattice sites that cannot be mapped onto each other by the lattice symmetry operations, may generally have different $l^{ph}$. 
In case of MnPS$_3$, the inequivalent spin-up and spin-down Mn atoms are found to have opposite $l^o$, see below. 
Similarly, Mo and S atoms in MoS$_2$ have opposite $l^o$ \cite{zhang2015chiral}.
The PCPs of all atoms in the unit cell must add up to zero. 
In multi-atom lattices, the individual $s_{z}^{\sigma}$ may be much smaller than unity. 
However, if we are sure that $s_{z}^{\sigma}$ is non-zero (i.e. above the numerical noise level), we set $l^s$ to $+1$ or $-1$, inheriting the sign of $s_{z}^{\sigma}$, as we know that $l^s$ must be quantized (see entries in Tables I and II).

In a honeycomb lattice, e.g. graphite, one obviously has two sublattices $A$ and $B$ having opposite $l^{ph}$. 
In MoS$_2$, $A$ and $B$ correspond to Mo and S sublattices (two S atoms share the same $xy$-coordinates). 
A multi-atom lattice allows for exploration of different assignments of $A$ and $B$ sublattices. 
In MnPS$_3$, we use two assignments of $A$ and $B$ sublattices when calculating the PCP: a) the two non-equivalent spin-up and spin-down Mn atoms, b) the different sulfur atoms, both shown in Fig.~\ref{structmnps3}.
More details of calculating the phonon polarization have been outlined in the Appendix.
 
\section{Computational details}

We consider a monolayer of MnPS$_3$  with lattice constants a= 5.945 $\textup{\AA}$, c=20 $\textup{\AA}$ with anti-ferromagnetic ordering where Mn$_1$, and Mn$_2$ are the non-equivalent atoms with opposite spins as shown in Fig.~\ref{structmnps3}. We perform a structural relaxation such that the total forces per atom are less than $10^{-5}$ Ryd/au. The monolayer of MnPS$_3$ has a hexagonal lattice structure with a D$_{3h}$ point group symmetry. In MnPS$_3$ while the non-magnetic structure has inversion symmetry the anti-ferromagnetic ordering breaks it. While the non-magnetic structure is metallic, anti-ferromagnetic ordering opens a band gap.

We calculate the electronic structure of this monolayer MnPS$_3$ using density functional theory (DFT+U) as implemented in the Quantum Espresso code (version 7.2)~\cite{giannozzi2017advanced} using the PBEsol functional~\cite{PeRo08} and ultra-soft pseudo-potentials~\cite{prandini2021standard}. The effects of spin-orbit coupling are included using fully relativistic pseudo-potential\cite{prandini2021standard}. We choose the Dudarev formulation of the DFT+U and  apply an on-site Coulomb repulsion of $U=3$ eV in the d-shells of the non-equivalent Mn atoms to include the static effects of correlations \cite{strasdas2023electronic,Dudarev1998PhysRevB}. We choose an energy cut-off of 69~Ryd, a k-mesh of $6 \times 6 \times 2$ \cite{MoPa76}. 

We calculate the phonon bandstructure using the finite displacement method with PHONOPY~\cite{phonopy} using a $(3 \times 3 \times 1)$ supercell and a k-mesh of $(3 \times 3 \times 1)$ in combination with the Quantum Espresso DFT+U calculations. We do not include the SOC effects while studying phonons since SOC plays almost no role and enhances computational costs. We investigate the displacement of atoms corresponding to the different phonon modes at the valley points and search for possible circular motion in atoms in any of these modes. 
We compute the PCP to quantify chirality using the formula in Eq.~\ref{pcp}.

\section{Results}
\label{sec:res}
\begin{figure*}
\hbox to\linewidth{\hfill%
\includegraphics[width=0.90\linewidth]{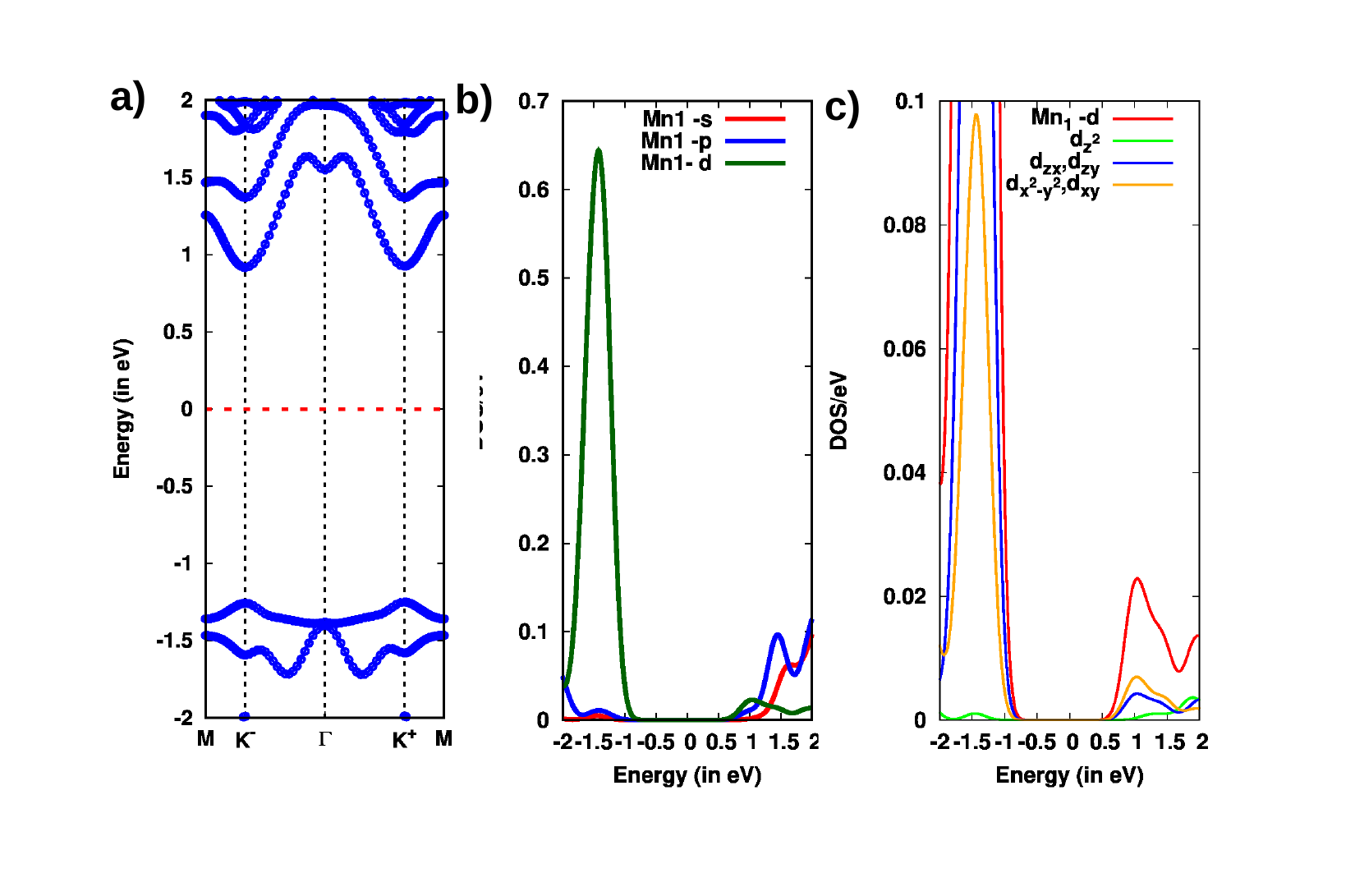}%
\hfill}
\caption{\label{bandsel}a) Electronic bandstructure of MnPS$_3$ with DFT+SOC+U, U=3eV, and (b-c) the partial electronic density of states near the band edges to identify the dominant character of the top valence and bottom conduction band.}
\end{figure*}

\subsection{Electronic band structure and valley polarization}

 In Fig.~\ref{bandsel}a) we show the electronic bandstructure of anti-ferromagnetic MnPS$_3$ calculated within DFT+U including the effects of spin-orbit coupling, and electronic correlations using $U=3$ eV. The inclusion of $U$ is necessary to obtain the dispersion across the valleys in agreement with Ref.~\onlinecite{li2013coupling}. 
 Having both the valance band maxima (VBM) and the conduction band minima (CBM) at the valley points $K$ makes the bandstructure look similar to MoS$_2$. 
 However, the combined $PT$-invariance of anti-ferromagnetic MnPS$_3$ results in Kramers degeneracy of the band energy, and consequently, unlike in TMDCs, no band splitting at the valleys is observed even in the presence of spin-orbit coupling. Nevertheless, according to Ref.~\onlinecite{li2013coupling} the orbital characters of the wave functions at the two valleys $K^+$ and $K^-$ are different due to the spin-orbit coupling effect, resulting in a spin-valley-dependent optical selection rule.  Thus, similar to the MoS$_2$ case, circularly polarized light can be used to build up valley polarization, and inter-valley transitions facilitated by electron-phonon coupling will produce phonons of a defined chirality. 
 We further see on analyzing the partial density of states at the band edges that the VBM is formed predominantly by  Mn $3d$ electrons, particularly the $d_{zx}$, $d_{zy}$, $d_{xy}$ and $d_{x^2-y^2}$ orbitals, as presented in Fig.~\ref{bandsel} b) and c). At the CBM, the same orbital character is present, but hybridized with sulfur $p$ orbitals, allowing for non-zero matrix elements for electric dipole transitions. 
 Our calculations are in overall agreement with previous experimental and theoretical literature \cite{li2013coupling,strasdas2023electronic}.

\subsection{Phonon bandstructure and PCP}
\begin{figure*}
\hbox to\linewidth{\hfill%
\includegraphics[width=0.90\linewidth]{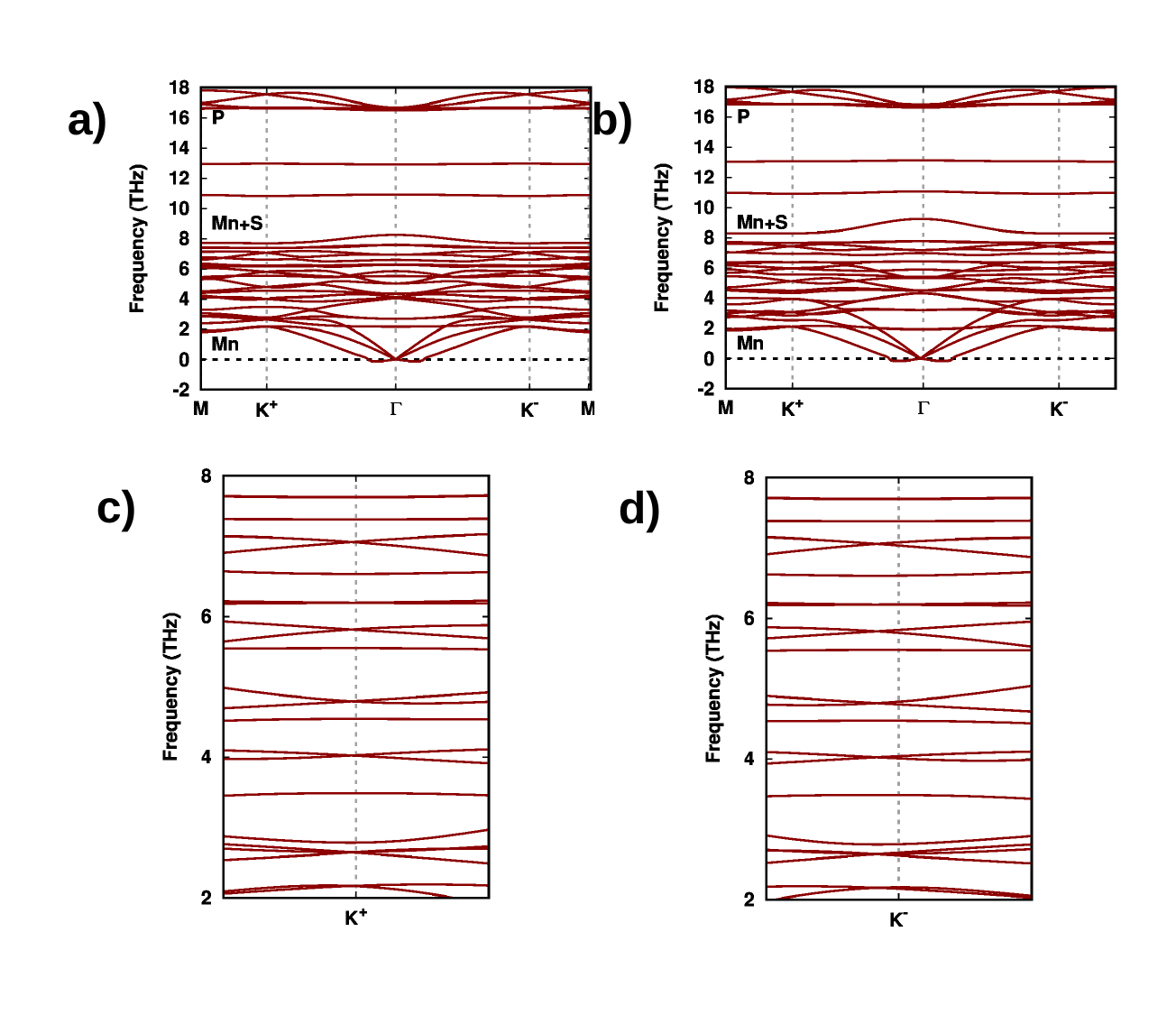}%
\hfill}
\caption{\label{phononbands} Phonon bandstructure in MnPS$_3$ with a) DFT-PBEsol and b) DFT+U c-d) magnified view of the Mn and S dominant bands around the valley points $K^{+}$ and $K^{-}$. }
\end{figure*}

In Fig.~\ref{phononbands} we show the phonon bandstructure with a) DFT-PBEsol and b) DFT+U where the $U= 3$eV is applied on the Mn d shell. The phonon modes at the $\Gamma$-point follow the D$_{3h}$ point group symmetry with $A$ and $E$ modes. Modes at the valley points $K^+$, $K^-$ as explicitly shown in Fig.~\ref{phononbands} (c-d) follow the C$_{3h}$ point group symmetry with the $A$, $E$, $E*$ modes. The doubly degenerate $E$ and $E*$ modes are the potential candidates for the chiral modes. The displacement vectors are real at the $\Gamma$-point while complex at the valley points, as expected. Our results agree with the calculated phonon frequencies in the existing literature \cite{yang2020electronic,hashemi2017vibrational,kargar2020phonon,zhang2022conductivity}.
We further note that frequencies are identical due to time-reversal symmetry for the $K^+$ and $K^-$ points. The lower frequencies are dominated by the Mn modes, intermediate frequencies are mixed Mn+S modes while the higher frequencies are the P modes, as marked in Fig.~\ref{phononbands}. On comparing the DFT-PBEsol and DFT+U bandstructures we see that the dispersion of the topmost Mn+S band gets stronger with the inclusion of U, particularly around the $\Gamma$-point.

\begin{figure*}
\hbox to\linewidth{\hfill%
\includegraphics[width=\linewidth]{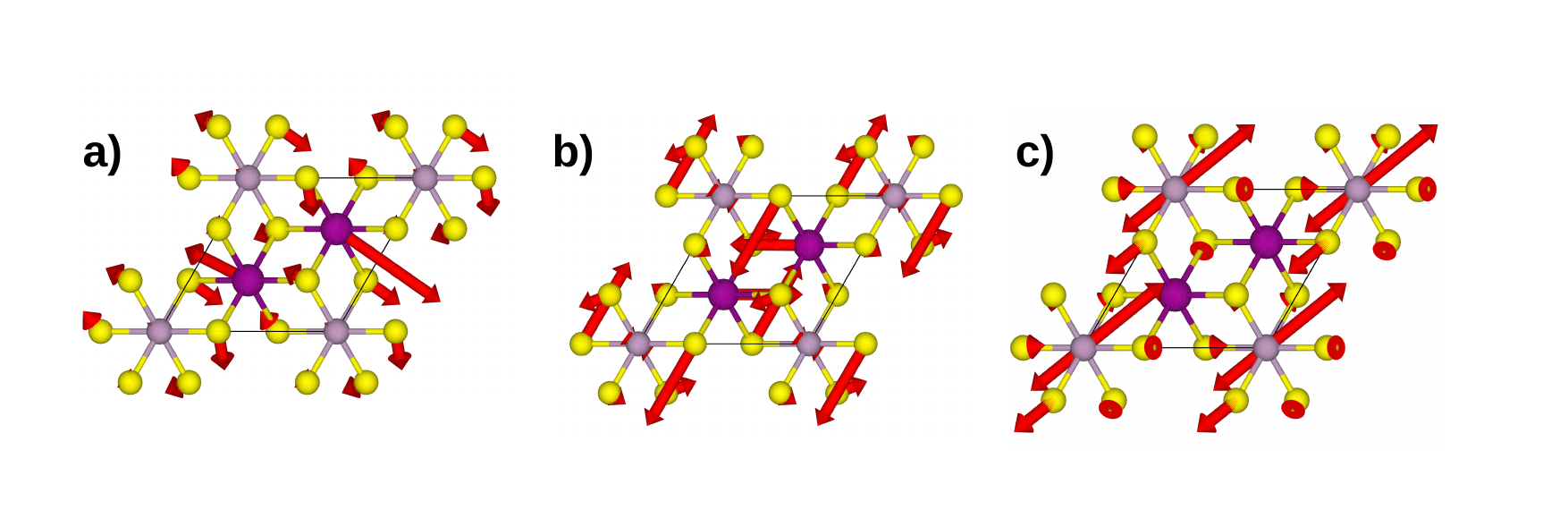}%
\hfill}
\caption{\label{chiraldisp} Displacement of atoms for a)5$^{th}$ mode b) 13$^{th}$ c) 28$^{th}$ mode. Circular motion of Mn (S) atoms are seen for the chiral modes 5 (13). Mode 28 is a non-chiral mode with dominant motion of the P atoms. Arrows indicate the displacement pattern of the respective atoms.}
\end{figure*}

We next investigate the displacement patterns of the different phonon modes in the valley points. We identify three types of displacement patterns as shown in  Fig.~\ref{chiraldisp}: a) Mn modes have circular motion (e.g. 5$^{th}$ mode), b) S modes have circular motions (e.g. 13 $^{th}$ mode), c) non-chiral P modes  
(e.g. 28$^{th}$ mode). The arrows indicate the direction of circular motions and we identify pairs of Mn and S atoms showing chirality, i.e., if one moves in the clockwise, the other in the anti-clockwise direction. This already indicates signatures of  chiral modes corresponding to the circular motion of Mn and S and non-chiral modes corresponding to the  
motion of P in MnPS$_3$. The inclusion of U does not change this pattern. 
However, it is cumbersome to quantitatively understand chirality when looking individually into the displacement pattern of the different modes, and hence we try to quantify the chirality of the various phonon modes calculating the PCP as defined in Eq.~\ref{pcp}.

First we discuss the results for the case where site A (B) are the two non-equivalent Mn$_{1}$ (Mn$_{2}$), as marked in Fig.~\ref{structmnps3}a). 


\begin{figure} [ht!]
\centering
\includegraphics[width=0.81\textwidth]{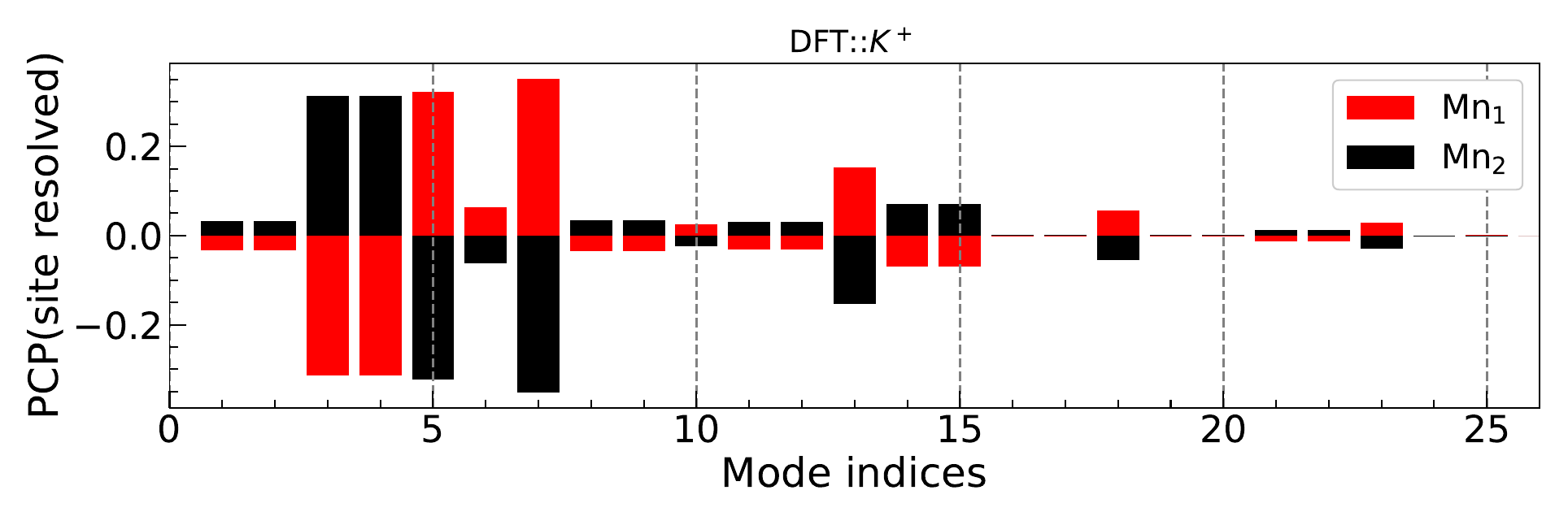} 
\qquad
\includegraphics[width=0.81\textwidth]{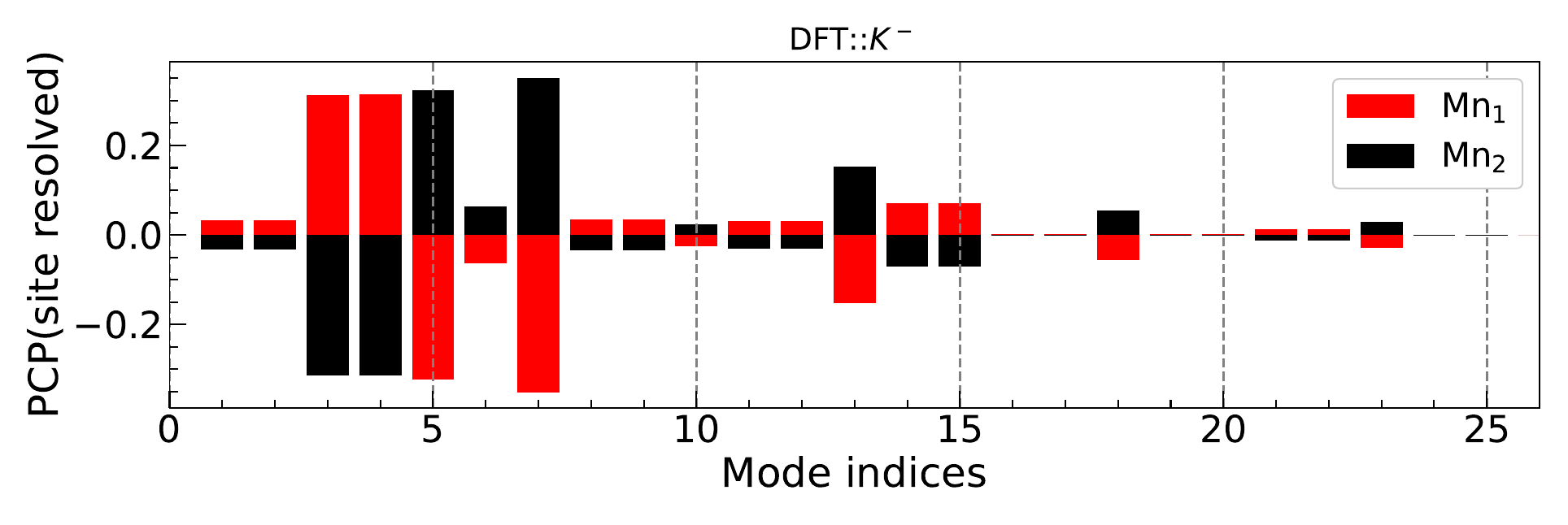} 
\caption{PCP for the different modes at the $K^+$ (upper panel) and $K^-$ (lower panel) using DFT-PBEsol.}
\label{pcpdft}
\end{figure}

\begin{figure} [ht!]
\centering
\includegraphics[width=0.81\textwidth]{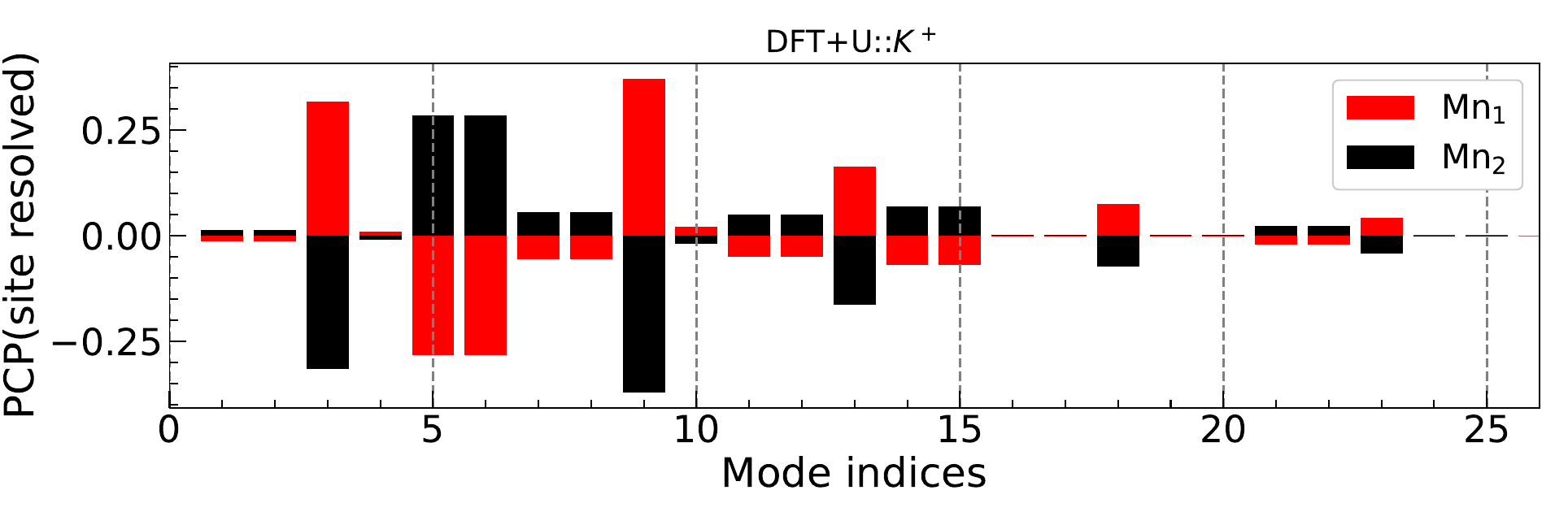} 
\qquad
\includegraphics[width=0.81\textwidth]{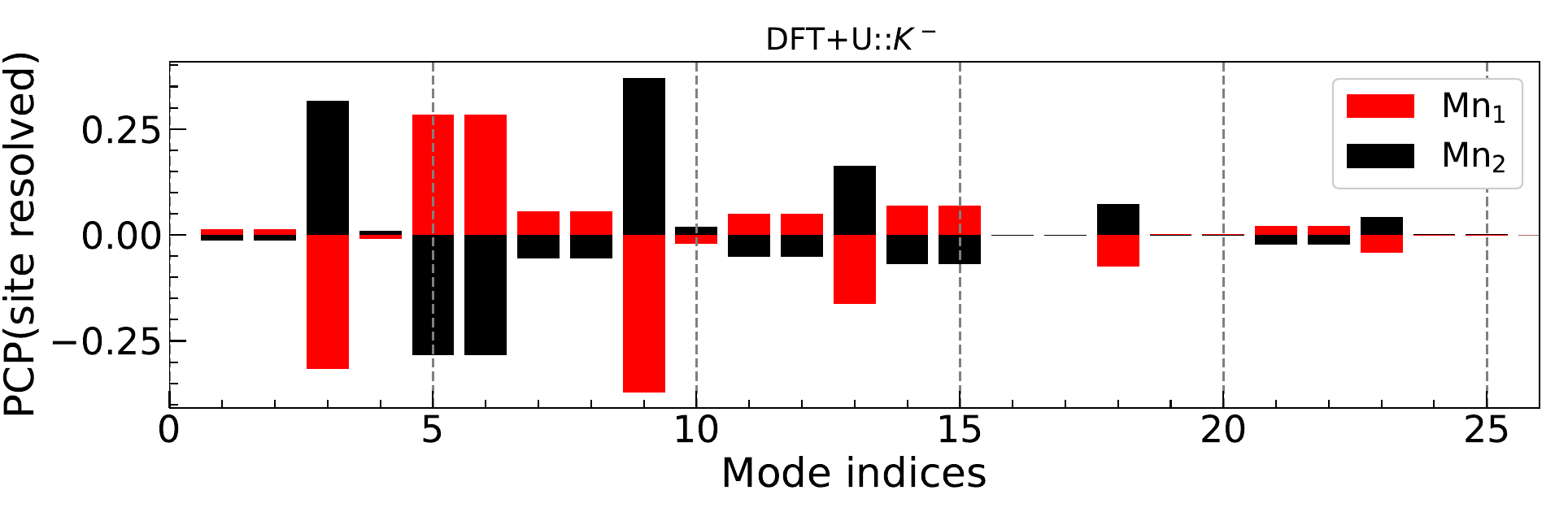} 
\caption{PCP for the different modes at the $K^+$ (upper panel) and $K^-$ (lower panel) using DFT+U.}
\label{pcpdftu}
\end{figure}
In Fig.~\ref{pcpdft} and Fig.~\ref{pcpdftu} we show the PCP for the different modes at high-symmetry points $K^{+}$ (upper panel) and $K^{-}$ (lower panel) for DFT-PBEsol and DFT+U, respectively. For $K^+$, the PCP due to site A (Mn$_{1}$) is shown in red and that due to site B (Mn$_{2}$) is shown in black. In both the cases we display  the chiral symmetry between the PCP of site A (red) and site B (black) for modes up to 25, since beyond it the PCP vanishes. We further see the time-reversal symmetry in the chiral modes between $K^+$ and $K^-$, i.e., the red and black bars are interchanged, showing a change of sign. Thus, the chiral symmetry is not broken with the inclusion of U.

\begin{table*}
\caption{\label{tab:chiral_dft}Selected chiral phonon modes ($\sigma$) from Fig.~\ref{pcpdft} and PCP larger that 0.05 in the K valley of MnPS$_3$ with phonon frequencies $\omega_{ph}$ (in THz). Mn$_{1}$ and Mn$_{2}$ are site A and B and thus the orbital angular momentum at site A and site B are $l^{o}_{Mn_{1}} =+1$ and $l^{o}_{Mn_{2}}= -1$, $l^{s}_{Mn_{1}}$ ($l^{s}_{Mn_{2}}$) are the circular polarization of Mn$_{1}$ (Mn$_{2}$). The phonon PAM $l^{ph}$ is given by $l^{o}_{Mn_{1}}+l^{s}_{Mn_{1}}$. 
We note that this addition must be done modulo 3 and can thus only yield 0,1 or $-1$.
}
\begin{ruledtabular}
\begin{tabular}{c|c|c|c|c|c|r}
 $\sigma$ & $\omega_{ph}$ & $s^{z}_{Mn_{1}}$ & $s^{z}_{Mn_{2}}$ & $l^{s}_{Mn_{1}}$ & $l^{s}_{Mn_{2}}$ & $l^{ph}_{Mn_{1}}$ \\
\hline
3  & 2.64  & $-0.3$ & 0.3 & $-1$ &+1 &0 \\
4& 2.64& $-0.3$ &+0.3 & $-1$ & +1 & 0 \\
5 & 2.65& 0.33& $-0.33$ &+1& $-1$ & $-1$ \\
6 & 2.78& 0.05& $-0.05$ &+1 & $-1$ & $-1$ \\
7 & 3.48&0.35 & $-0.35$ & +1 & $-1$ & $-1$ \\
13 &5.55&0.15& $-0.015$ &+1 & $-1$ &$-1$ \\
14 &5.95& $-0.05$ &0.05 & $-1$ &+1 & 0\\
15 &5.95& $-0.05$ &0.05 & $-1$ &+1 & 0\\
18 & 6.68& 0.05 & $-0.05$ &+1 & $-1$ & $-1$\\
\end{tabular}
\end{ruledtabular}
\end{table*}

The explicit values of the frequencies, PCP corresponding to the chiral modes, pseudo-angular momentum (PAM) and phonon polarization corresponding to DFT-PBEsol and DFT+U are presented in Table~\ref{tab:chiral_dft} and Table~\ref{tab:chiral_dft_u}, respectively, in line with the MoS$_2$ example presented in  Ref.~\onlinecite{zhang2015chiral}. 
In the DFT+U method the phonons generally come out harder than in DFT-PBEsol. However, different modes are affected to a different extent. This leads to a re-indexing of the modes as presented in Table~\ref{tab:chiral_dft_u}. Especially the acoustic modes become significantly harder (up to 40\%), whereas the highest modes with major contribution from the phosphorus atoms are less affected (around 10\%). The modes with the highest PCP are the acoustic modes (number $\sigma=5$ and 7 in DFT-PBEsol, number $\sigma=3$ and 7 in DFT+U), displaying the same sign of $s_z^\sigma$ for a given Mn sublattice. 
In addition, there is a pair of degenerate modes (number 3,4 in DFT-PBEsol, number 5,6 in DFT+U) that share the same $s_z^\sigma$. However, the sign of their PCP is opposite to the one found in the acoustic modes.

A second group of modes shows much smaller PCP values.
Inspecting the motion of the atoms shows that the displacement in these modes happens mostly at the S atoms, and the Mn atoms (that were used to calculate the PCP) perform minor displacements. The PCP of these modes is in the range of 0.05 to 0.15.

\begin{table*}
\caption{\label{tab:chiral_dft_u} Selected cases showing the re-indexing of modes from $\sigma_{1}\rightarrow \sigma_{2}$ where $\sigma_{1}$ corresponds to the DFT-PBEsol mode in  Table~\ref{tab:chiral_dft} and $\sigma_{2}$ corresponds to the re-indexed mode of DFT+U.}
\begin{ruledtabular}
\begin{tabular}{c|c|c|c|c|c|r}
 $\sigma$&$\omega_{ph}$&$s^{z}_{Mn_{1}}$& $s^{z}_{Mn_{2}}$&$l^{s}_{Mn_{1}}$ &$l^{s}_{Mn_{2}}$&$l^{ph}_{Mn_{1}}$ \\
\hline
$5\rightarrow 3$  & 2.53  &0.33 & $-0.33$ &+1 & $-1$ & $-1$ \\
$3 \rightarrow 5$ & 3.07 & $-0.25$ & 0.25 & $-1$ & +1& 0 \\
$4 \rightarrow 6$ & 3.07& $-0.25$ & 0.25 & $-1$ & +1 & 0 \\
$7 \rightarrow 9$ & 4.33 & 0.38 & $-0.38$ & +1 & $-1$ & $-1$ \\
\end{tabular}
\end{ruledtabular}
\end{table*}

Finally, the modes above 6 meV are found to have vanishing chirality, with the exception of mode 18 and 23 having PCPs of about 0.05. Notably, mode 23 is the highest-lying mode of the Mn+S mode spectrum; all higher modes have significant P contributions. Their motion is essentially decoupled from the other modes, and hence their PCP vanishes.

PAM has contributions from the local (spin) part $l_{s}$ and the non-local (orbital) part $l_{o}$. $l^{o}_{Mn_{1}}=1$ and $l^{o}_{Mn_{2}} = -1$ as site A and B are in opposite phases. Thus $l^{s}_{Mn_{1}}$ can be either +1 or $-1$.  Since $l^{ph}$ is $l^{o}_{Mn_{1}}+l^{s}_{Mn_{1}}$, it can therefore be $-1$ ($1+1=2\equiv -1$) or 0 for the chiral modes listed. We note that this addition must be done modulo 3 and can thus only yield 0,1 or $-1$. 

A similar analysis was carried out for the assignment of  $A$ and $B$ to the sulfur sublattices, as shown in Fig.~\ref{structmnps3}b). In this case, no chiral behavior is seen.

\section{Conclusions and outlook}

Signatures of chiral phonon modes in a monolayer of the anti-ferromagnetic semiconductor MnPS$_3$ with hexagonal lattice structure have been found and characterized by their phonon angular momentum. While chiral phonon modes had previously been theoretically studied in non-magnetic TMDCs, such as MoS$_2$, WS$_2$ and WSe$_2$, signatures of chiral phonon modes in a honeycomb lattice with anti-ferromagnetic ordering is presented here for the first time to our knowledge.

It would be interesting to see if the chiral phonon modes can be experimentally observed in this material using ultra-fast electron diffraction or other innovative experimental techniques, as discussed in Ref.s~\onlinecite{britt2023ultrafast,gerbig2025polarization,caruso20252025}. Since we calculate phonons at the valley points rather than the $\Gamma$-point, Raman spectroscopy is not a suitable tool for this purpose. 
In particular, experiments sensitive to local magnetic moments might be able to detect the alternating sign of the PCP predicted by our calculations. 

As future work, it is interesting to calculate the electron-phonon coupling of the chiral modes to the electrons in the valleys to better understand the driving of the chiral modes by circularly polarized light. Moreover, phonon chirality might influence the thermal properties, e.g. thermal conductivity \cite{zhang2022conductivity}.
It is further interesting to see how the chiral phonon modes are affected if we go from a monolayer to multi-layers of MnPS$_3$. While bilayers have an inversion center between the layers, for trilayers this depends on the stacking.
Finally, one could extend our studies to other anti-ferromagnets of this family of materials having similar hexagonal lattice structure, e.g. FePS$_3$ and NiPS$_3$.

\section*{Acknowledgments}
This work is funded by the Deutsche Forschungsgemeinschaft (DFG, German Research Foundation) – Project-ID 278162697– SFB 1242. 
The authors gratefully acknowledge the computing time granted by the Center for Computational Sciences and Simulation (CCSS) of the University of Duisburg-Essen and provided on the supercomputer magnitUDE (DFG Grant No. INST 20876/209-1 FUGG and INST 20876/243-1 FUGG) at the Zentrum f{\"u}r Informations- und Mediendienste (ZIM). We thank L. Zhang for fruitful discussions.

\appendix
\section{Formalism of phonon circular polarization (PCP)}
We briefly sketch the concept of phonon circular polarization following Ref.~\onlinecite{zhang2015chiral}. 
The phonon eigenmodes in an $n$-atom basis is given by $\epsilon = (x_{1}, y_{1}, \ldots x_{n, }y_{n})^{T}$. The vibration in the $xy$-plane yields angular momentum in the $z$-direction. Next we define a new basis with right and left circular polarization operators as follows:
\begin{equation}
\begin{array}{cccc}
\ket{R_1} \equiv \frac{1}{\sqrt{2}} 
\begin{pmatrix}
1 \\ \mathrm{i} \\ \vdots \\ 0
\end{pmatrix},
& \cdots &
\cdots
& \ket{R_n} \equiv \frac{1}{\sqrt{2}} 
\begin{pmatrix}
0 \\ \vdots \\ 1 \\ \mathrm{i}
\end{pmatrix},
\end{array}
\tag{A1}
\end{equation}

\begin{equation}
\begin{array}{cccc}
\ket{L_1} \equiv \frac{1}{\sqrt{2}} 
\begin{pmatrix}
1 \\ -\mathrm{i} \\ \vdots \\ 0
\end{pmatrix},
& \cdots &
\cdots
& \ket{L_n} \equiv \frac{1}{\sqrt{2}} 
\begin{pmatrix}
0 \\ \vdots \\ 1 \\ -\mathrm{i}
\end{pmatrix}\, .
\end{array}
\tag{A2}
\end{equation}
We can define the phonon mode $\epsilon$ in the R-L basis with $\alpha$ as the atomic index as 
\begin{equation}
 \epsilon= \sum_{\alpha=1}^{n} \left[\epsilon_{R_{\alpha}} \ket{{R_{\alpha}}}+\epsilon_{L_{\alpha}} \ket{{L_{\alpha}}}\right],
 \tag{A3}
\end{equation}
where
\begin{equation}
 \epsilon_{R_{\alpha}}=\bra{R_{\alpha}}\ket{\epsilon}=\frac{1}{\sqrt{2}}\left(x_{\alpha}-iy_{\alpha}\right),
 \tag{A4}
\end{equation}

\begin{equation}
 \epsilon_{L_{\alpha}}=\bra{L_{\alpha}}\ket{\epsilon}=\frac{1}{\sqrt{2}}\left(x_{\alpha}+iy_{\alpha}\right)\, .
 \tag{A5}
\end{equation}
Due to the completeness relation, we have
\begin{equation}
 \epsilon^{\dag}\epsilon =\sum_{\alpha}  |\epsilon_{R_{\alpha}}|^{2}+|\epsilon_{L_{\alpha}}|^{2}=1\, .
  \tag{A6}
\end{equation}
 We define  the phonon polarization operator as 
 \begin{equation}
  \hat{s}_{z}=\sum_{\alpha=1}^{n}(\ket{R_{\alpha}}\bra{R_{\alpha}}-\ket{L_{\alpha}}\bra{L_{\alpha}})\, .
   \tag{A7}
 \end{equation}
 In case of a hexagonal lattice each sublattice can have the polarization 
 $\sum_\alpha \epsilon_\alpha ^{\dag}\hat{s}_{z}\epsilon_\alpha$ where $\alpha$ runs only over the sublattice sites. 
 The relation with phonon angular momentum is worked out in Section~\ref{sec:matmethod} of the main text.
%

The mode vectors $\epsilon$ entering in Eq.~\ref{eq:2nd_quanti} satisfy an eigen value problem, 
\begin{equation}
 \boldsymbol{D}(\mathbf{k}) \; \epsilon(\mathbf {k},\sigma) =\omega^{2}_{\mathbf{k,\sigma}} \; \epsilon(\mathbf {k},\sigma)
\end{equation}
where $\boldsymbol{D}$  is the dynamical matrix with elements 
\begin{equation}
 D_{\alpha \alpha '}(\mathbf{k}) = \sum_{R_{t'}-R_{t}} \frac{\mathbf{K}_{t\alpha, t'\alpha'}}{\sqrt{m_{\alpha} m_{\alpha '}}} e^{i (\mathbf{R}_{t'}-\mathbf{R}_{t})\cdot \mathbf{k}} \, .
\end{equation}
$D_{\alpha \alpha '}$ is $d n \times d n$ square matrix, where $d$ is the dimension of the system 
and $n$ the number of atoms in the unit cell. $\mathbf{K}_{t\alpha,t'\alpha'}$ is the spring constant matrix between the $\alpha$-th atom in the $t$-th unit cell and $\alpha'$-th atom in the $t'$-th unit cell. From the dynamical matrix one can obtain orthonormal eigen vectors as $\epsilon^{\dag} (\mathbf{k}\sigma)$ and $\epsilon (\mathbf{k}\sigma')$ satisfying the completeness relation.\\
Using these orthonormal eigen vectors once can obtain the phonon circular polarization defined in Eq.~\ref{pcp} and presented in this paper.

\bibliography{mnps3}
\end{document}